\documentclass[10pt,twoside,twocolumn,amsmath,amssymb,showpacs]{revtex4}
\usepackage[utf8]{inputenc}
\usepackage{color}
\usepackage{bm}
\usepackage{amsmath}
\usepackage{amssymb}
\usepackage{graphicx}
\usepackage{wasysym}

\makeatletter
\@ifundefined{textcolor}{}
{%
 \definecolor{BLACK}{gray}{0}
 \definecolor{WHITE}{gray}{1}
 \definecolor{RED}{rgb}{1,0,0}
 \definecolor{GREEN}{rgb}{0,1,0}
 \definecolor{BLUE}{rgb}{0,0,1}
 \definecolor{CYAN}{cmyk}{1,0,0,0}
 \definecolor{MAGENTA}{cmyk}{0,1,0,0}
 \definecolor{YELLOW}{cmyk}{0,0,1,0}
}

\usepackage{mathrsfs}

\usepackage{amsfonts}
\usepackage{dcolumn}
\usepackage{bm}

\makeatother

\begin{document}

\title{Emergence of Living Chiral Superlattice from \textit{Biased-Active} Particles}

\author{Yongliang Gou, Huijun Jiang$^*$, Zhonghuai Hou}

\thanks{Corresponding Author: hzhlj@ustc.edu.cn; hjjiang3@ustc.edu.cn}

\affiliation{Department of Chemical Physics \& Hefei National
Laboratory for Physical Sciences at Microscales, iChEM, University
of Science and Technology of China, Hefei, Anhui 230026, China}

\date{\today}
\begin{abstract}
We introduce for the first time a general model of \textit{biased-active} particles, where the direction of the active force has a biased angle from the principle orientation of the anisotropic interaction between particles. We find that a highly ordered living superlattice consisting of small clusters with dynamic chirality emerges in a mixture of such biased-active particles and passive particles. We show that the biased-propulsion-induced instability of active-active particle pairs and rotating of active-passive particle pairs are the very reason for the superlattice formation. In addition, a biased-angle-dependent optimal active force is most favorable for both the long-range order and global dynamical chirality of the system. Our results demonstrate the proposed \textit{biased-active} particle providing a great opportunity to explore a variety of new fascinating collective behaviors beyond conventional active particles.
\end{abstract}

\maketitle
Since active systems can be driven far from equilibrium\cite{ramaswamy2010mechanics} by continuously consuming energy supplied internally or externally, understanding collective behaviors of such systems is of great importance for revealing the mystery of living systems and further for manufacturing smart materials\cite{zhang2017active}. To date, many research interests have been paid on the effect of active motion on the dynamics, and a great array of collective behaviors that cannot be manifested in equilibrium systems have been reported, such as motility-induced phase separation\cite{p2,p1}, anomalous density fluctuations\cite{p3,p4}, and spontaneous flow\cite{p5}.

Recently, it is recognized that designing complex interactions between active particles rather than simply changing the active force is very important for guiding the formation of collective behaviors\cite{zhang2017active}. Chemically synthesized Janus particles is one of the examples, where self-propulsion may arise from non-uniform properties of the Janus particles\cite{p10,p11,p12,p13} while the interaction between these particles can be strongly anisotropic\cite{Tarazona}. By controlling the amplitude of anisotropic interactions, S.Granick and coworkers\cite{p6,p7,zhang2016directed} reported several interesting new dynamic phase states such as rotating pinwheels. It should be a significant step forward in active-particle designing if new methods beyond conventional ones are proposed.

In this Letter, we propose a conceptually new design of active particles focusing on the correlation between active motion and  anisotropic interaction, namely, \textit{biased-active} particles where the direction $\mathbf{n}$ of active force has a biased angle $\theta$ from the principle orientation $\mathbf{q}$ of the anisotropic interaction (Fig.1). The variation of $\theta$ offers a rich design space for dynamic self-assembly, providing a great opportunity to explore a variety of new fascinating collective behaviors beyond conventional active particles. As an example, we report the emergence of a striking superlattice structure with dynamic chiral clusters (DCCs) in a mixture of such biased-active particles and passive particles for $\theta$ larger than some threshold values. We find that such biased propulsion may on one hand lead to instability of active particle (AP) clusters formed due to anisotropic interaction, and on the other hand induce the rotation motion of an AP around the passive particle (PP) it attached.
As a consequence, many ordered hexagonal DCCs, each with six APs rotating around a PP, are formed. These DCCs may finally organize into a superlattice with long range order and hexagonal symmetry which would not be observed neither in the counterpart equilibrium system nor in the conventional active particle system.

\begin{figure}

\begin{centering}
\includegraphics[width=0.8\columnwidth]{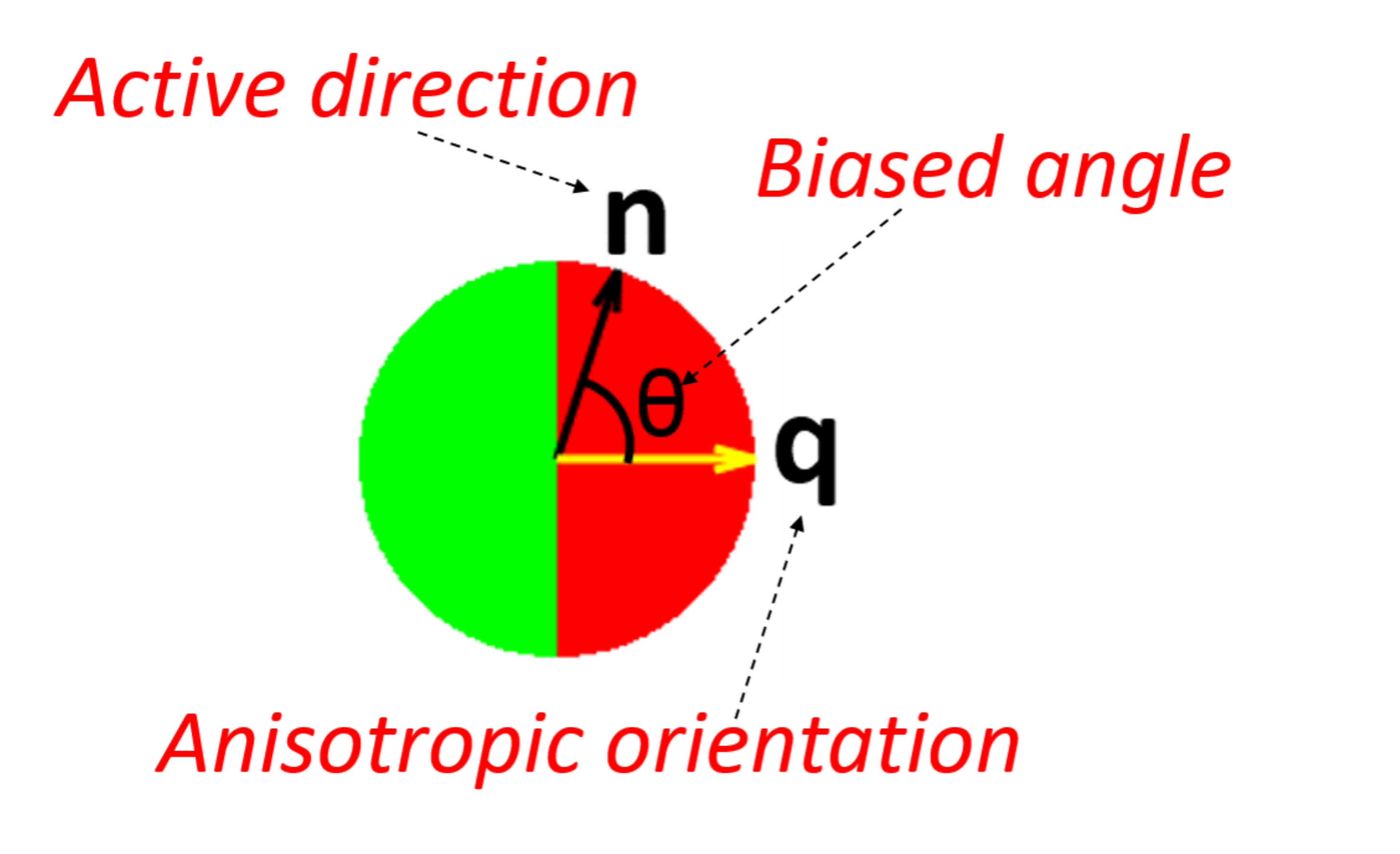}
\end{centering}
\caption{Schematic of the biased-active particle with a biased-angle $\theta$
between active direction $\mathbf{n}$ and the anisotropic interaction orientation
$\mathbf{q}$. The red side is attractive and the green is repulsive. \label{Fig_Model}}
\end{figure}

The system contains $N_{a}$ APs and $N_{p}$ PPs
of the same diameter $\sigma$. Each AP is of Janus type containing
two half spheres which allows us to define an orientation unit vector
$\mathbf{q}$, pointing to the face (red side) as shown in Fig.1.
Besides, AP is also subjected to a self-propulsion force with amplitude
$F_{a}$ along a direction given by a unit vector $\mathbf{n}$
with a biased angle $\theta$ from $\mathbf{q}$. For a pair of APs $i$ and $j$, the interaction
$U_{ij}$ contains two terms,

\begin{equation}
U_{ij}(\textbf{r}_{ij},\textbf{q}_{i},\textbf{q}_{j})=U_{WCA}(r_{ij})+U_{AN}(\textbf{r}_{ij},\textbf{q}_{i},\textbf{q}_{j}),
\end{equation}
where $\mathbf{r}_{ij}=\mathbf{r}_{j}-\mathbf{r}_{i}$ is the vector
pointing from particles $i$ to $j$, $r_{ij}=\left|\mathbf{r}_{ij}\right|$
is the corresponding distance. The first term $U_{WCA}\left(r_{ij}\right)$
denotes an isotropic excluded volume interaction given by the WCA
potential\cite{WCA}, $U_{WCA}(r_{ij})=4\epsilon\left[(\frac{\sigma}{r_{ij}})^{12}-
(\frac{\sigma}{r_{ij}})^{6}+\frac{1}{4}\right]$
if $r_{ij}<2^{1/6}\sigma$, and zero otherwise,
with $\epsilon$ the interaction strength. The second term $U_{AN}\left(\mathbf{r}_{ij},\mathbf{q}_{i},\mathbf{q}_{j}\right)$
is the anisotropic Yukawa interaction potential given by \cite{Yuk,Yukawa}
\begin{equation}
U_{AN}(\textbf{r}_{ij},\textbf{q}_{i},\textbf{q}_{j})=\frac{C\exp[-\lambda(r_{ij}-\sigma)]}{r^2_{ij}}(\textbf{q}_i-\textbf{q}_j)\cdot\textbf{\ensuremath{\textbf{r}}}_{ji}
\label{eq:U_AN},
\end{equation}
where $C$ denotes the interaction strength and $\lambda^{-1}$ gives the
corresponding screen length. According to this anisotropic interaction,
two APs $i$ and $j$ attract each other most strongly if their orientation
vectors $\mathbf{q}_{i}$ and $\mathbf{q}_{j}$ pointing to each other,
i.e., in a face-to-face configuration, while they repel each other
most strongly in back-to-back position. The interactions $U_{ij}$
between any two PPs are isotropic and just given by the WCA potential with same parameters as those for APs.
For the interactions between a pair of PP and AP, we assume that PP
only attracts the face side of AP and has no interactions with the
back side. The interaction potential is also given by Eq.(\ref{eq:U_AN}),
but with a stronger strength $C'$ for active-passive pairs than $C$
for active-active pairs.

The evolution equation governing the dynamics of
$\mathbf{r}_{i}$ ($i=1,2,...,N$ with $N=N_a+N_p$) is then given by

\begin{equation}
\dot{\textbf{r}}_{i}=\frac{D_{t}}{k_{B}T}\left(\sum_{j\neq i}^{N}-\frac{\partial U_{ij}}{\partial\mathbf{r}_{i}}+F_{a}\textbf{n}_{i}\right)+\mathrm{\bm{\xi}}_{i}\label{eq:Ri_Motion},
\end{equation}
where $k_{B}$ denotes the Boltzmann constant, $T$ is the temperature,
and $\bm{\xi}_{i}$ is the thermal fluctuation satisfying the fluctuation-dissipation
relationship $\left\langle \bm{\xi}_{i}(t)\bm{\xi}_{j}(t')\right\rangle =2D_{t}\bm{1}\delta_{ij}\delta(t-t')$
with $D_{t}$ the translational diffusion coefficient and $\mathbf{1}$
the unit tensor. $F_{a}$ is the amplitude of active force and is set to be zero for PPs.

For an AP, the direction of active force $\mathbf{n}_{i}$
(and thus $\mathbf{q}_{i}$) changes via random rotational diffusion.
In addition, the anisotropic interaction also exerts a torque on the
particle and thus leads to the change of particle orientation.
Therefore, the dynamic equation of $\mathbf{q}_{i}$ can be written
as,

\begin{equation}
\dot{\textbf{q}}_{i}=-\frac{D_{r}}{k_{B}T}\sum_{j\neq i}^{N}\frac{\partial U_{AN}(\mathbf{r}_{ij},\mathbf{q}_{i},\mathbf{q}_{j})}{\partial\mathbf{q}_{i}}+\bm{\eta}_{i}\times\textbf{q}_{i}\label{eq:ni_Motion},
\end{equation}
where the first term describes the torque exerted by the aforementioned
anisotropic interaction to particle $i$, and $\bm{\eta}_{i}$
is the rotational fluctuation of the active direction satisfying $\left\langle \bm{\eta}_{i}(t)\bm{\eta}_{j}(t')\right\rangle =2D_{r}\bm{1}\delta_{ij}\delta(t-t')$
with the rotational diffusion coefficient $D_{r}=3D_{t}/\sigma^{2}$.

Simulations are performed in a $L\times L$ two
dimensional square box with periodic boundary conditions. $\sigma$,
$k_{B}T$, and $\tau=\sigma^{2}/(10D_{t})$ are chosen as the dimensionless
units for length, energy and time, respectively. We fix $N=2100$
with $N_{a}/N_{p}=6$, $\varepsilon=1.0$, $C=3.0$, $\lambda=3\sigma^{-1}$
during the simulations if not otherwise stated. The box length is
$L=60$ corresponding to a packing fraction $\text{\ensuremath{\phi}=0.46}$,
and the simulation time step is $\Delta t=10^{-4}\tau$. We assume
a stronger interaction among passive-active pairs than active-active
ones by setting $C'=2C$ in the current work. The active force $F_{a}$
and biased angle $\theta$ are chosen as variable parameters. All
simulations start from random initial conditions and run for enough
long time to ensure the system has reached a stationary state.

\begin{figure}
\begin{centering}
\includegraphics[width=1.1\columnwidth]{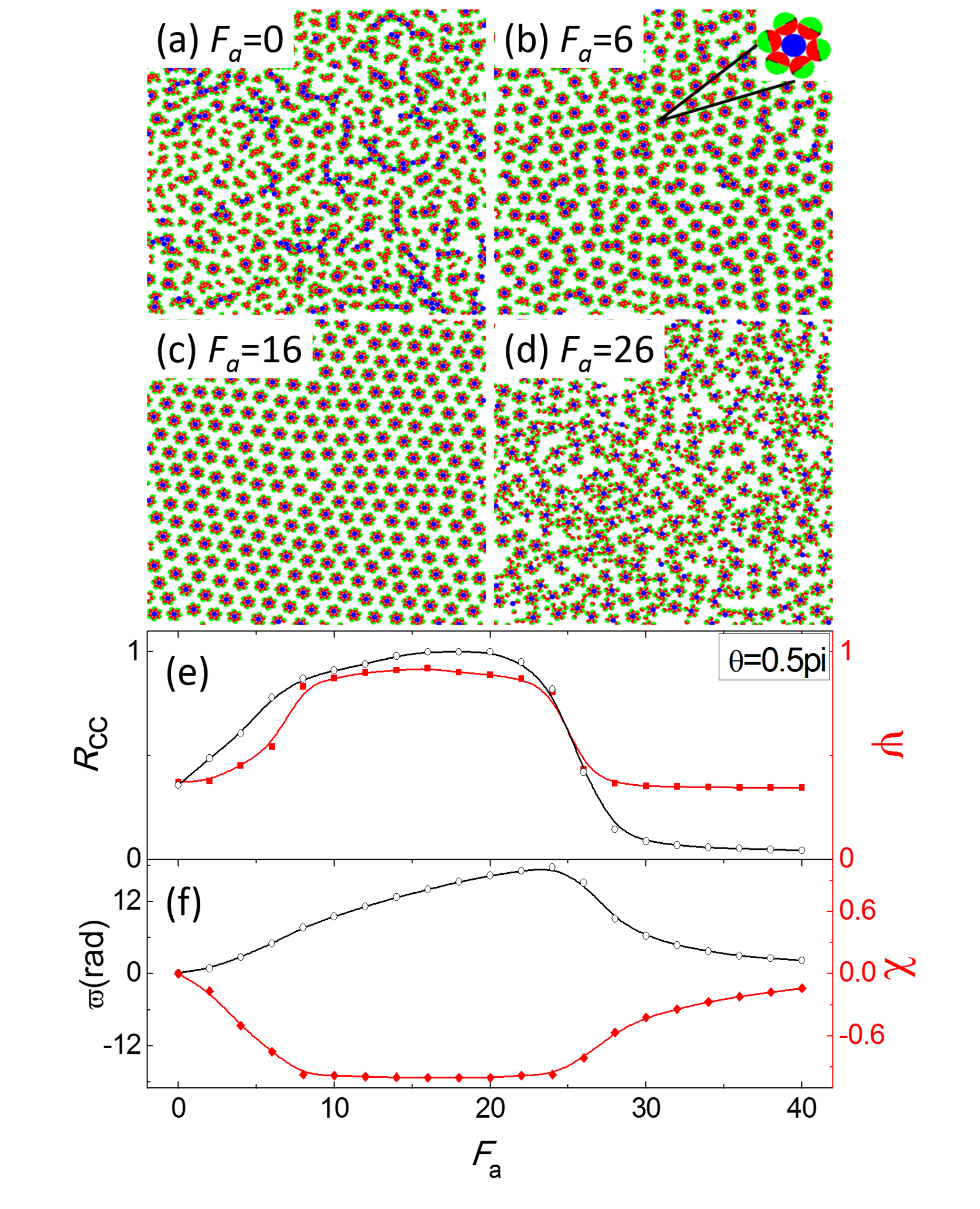}
\end{centering}
\caption{(a)-(d) Typical snapshots for active forces $F_{a}=0$, 6, 16 and 26,
respectively, with $\theta=0.5\pi$. The inset in (b) shows zoom-in of the dynamic chiral
cluster (DCC) with six active particles (red-green) surrounding a passive
one (blue). (e) The fraction of DCCs, $R_{cc}$ (left), and the global order parameter, $\Psi$
(right), as functions of $F_{a}$. (f) The average rotation speed $\bar{\omega}$
of DCCs (left) and the global dynamic chirality $\chi$ as functions of $F_{a}$. \label{Fig_Theta0.5}
}
\end{figure}

Firstly, we consider a typical angle $\theta=\text{\ensuremath{\pi}/2}$,
where the active force direction $\mathbf{n}_{i}$ for particle
$i$ ($i=1,...,N$) is perpendicular to the principle direction of anisotropic interaction
$\mathbf{q}_{i}$. Very interestingly, it is observed that a highly ordered
superlattice consisting of many living chiral clusters emerges spontaneously
if the active force is within a certain appropriate range. In Fig.\ref{Fig_Theta0.5}(a)-(d),
the typical snapshots of the system are depicted
for different active forces $F_{a}=0$, 6, 16, and 26, respectively.
Without activity ($F_{a}=0$), the particles tend to form small clusters
with the face sides attracted together due to the anisotropic
interaction, or clusters with several APs attached to a PP due to
attractions between AP-PP pairs (Fig.\ref{Fig_Theta0.5}(a)). One can also find few hexagonal
clusters wherein one PP is surrounded by six APs, which are quite
ordered in short range, nevertheless, the whole system is disordered
in long range. Note that this disordered state is quite stable with
respect to thermal noises $\bm{\xi}$ in the system. For a small active
force, say $F_{a}=6$ as shown in Fig.\ref{Fig_Theta0.5}(b), much more
clusters with one PP surrounded by six APs (the inset in Fig.\ref{Fig_Theta0.5}(b)) emerge. Interestingly,
the APs rotate clockwise around the central PP continuously, demonstrating
a novel type of \textit{dynamic chirality}.
Some long-range order already appears in this state, nevertheless,
there are still many non-hexagon clusters remaining in the system.
For an appropriate level of activity as shown in Fig.\ref{Fig_Theta0.5}(c)
for $F_{a}=16$ , remarkably, a perfectly ordered superlattice emerges
with hexagon structures in both long and short ranges. In this superlattice
state, each PP is accompanied by six APs to form a hexagon cluster,
and all the $N_p$ clusters rotate in the same clockwise direction.
If the active force is too large, however, such ordered structure
is destroyed again as shown in Fig.\ref{Fig_Theta0.5}(d) for $F_{a}=26$,
wherein the long-range order is lost and many hexagon clusters are
broken. Supplemental movies are available for these $F_a$.

Clearly, the living one-plus-six dynamic chiral clusters (DCCs) play important roles in the system's collective behaviors.
With the increase of active force $F_{a}$, we find that the fraction
of DCCs (left axis in Fig.\ref{Fig_Theta0.5}(e)) undergoes a maximum value,
where $R_{cc}=N_{cc}/N_{p}$ with $N_{cc}$ the number of DCCs. For small
or large $F_{a}$, $R_{cc}$ is small (but not zero), while it reaches
nearly 1.0 within an intermediate range of $F_{a}$. To characterize
the long-range order of the system, we measure the global
order by the parameter $\Psi=N_{p}^{-1}\sum_{j=1}^{N_p}\overline{q_6^j}$
for the lattice formed by PPs, where the overbar denotes averaging
over time. $q_6^j=\frac{1}{6}\sum_{k\in\mathcal{N}\left(j\right)}\exp\left(i6\theta_{kj}\right)$
denotes the local order parameter for the $j$th PP with $k\in\mathcal{N}\left(j\right)$
running over its 6 nearest PP neighbors and $\theta_{kj}$ the angle
between an arbitrary axis and $\mathbf{r}_{kj}$. Fig.\ref{Fig_Theta0.5}(e)
also shows the dependence of $\Psi$ (right axis) on $F_{a}$, wherein
a clear-cut maximum can again be observed, i.e., $\Psi$ increases from
a relatively small value at $F_{a}=0$, to a value close to 0.9 for
intermediate values of $F_{a}$ corresponding to a very ordered superlattice
state, and then reduces sharply to a small value again for $F_a\apprge30$.
Clearly, the fraction of DCCs $R_{cc}$ is highly correlated with
the global order parameter $\Psi$. Such findings clearly demonstrate
that an optimal level of particle activity drives the formation of
the highly ordered superlattice.

\begin{figure}
\begin{centering}
\includegraphics[width=1.0\columnwidth]{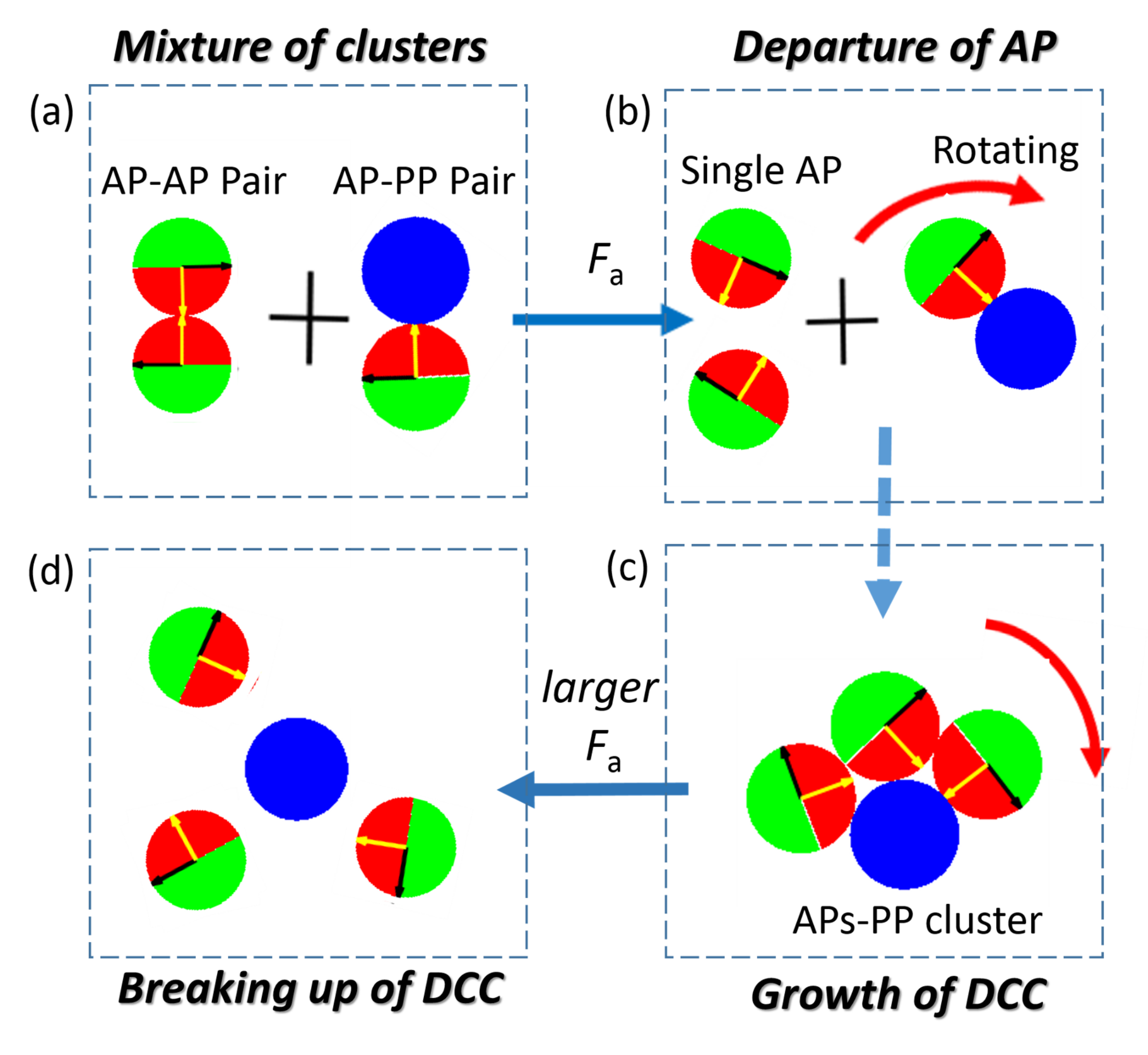}
\end{centering}
\caption{Schematic of the formation and breaking mechanism of DCC. The red arrows are the rotation
directions of AP–PP pairs. Taking the $\theta=0.5\pi$ as an example, (a) both of the AP–AP pairs and the AP–PP
pairs are formed for $F_{a}=0$. (b) Increase $F_{a}$ breaks the AP–AP pair into single APs, and the particle pairs start to rotate. (c) Single APs are attracted by PPs due to the stronger anisotropic interaction between AP-PP pairs. (d) For large enough $F_{a}$, the DCC also falls apart. \label{Fig_Mechanism}
}
\end{figure}

Another nontrivial feature of the superlattice as
shown in Fig.\ref{Fig_Theta0.5}(c) is the rotation of DCCs, i.e.,
the peripheral six APs rotate clockwise around the central PP. Interestingly, the average rotation speed $\bar{\omega}$
(calculated for APs stably rotated around the PP they attach) also
depends non-monotonically on the active force as shown in Fig.\ref{Fig_Theta0.5}(f)
(left axis). For $F_{a}=0$, although some hexagon clusters also exist,
they just randomly swing and $\bar{\omega}$ is nearly zero. With
increase of $F_{a}$, the rotation speed $\bar{\omega}$ also increases
till it reaches a maximum value $F_{a}\sim24$, after which $\bar{\omega}$
decreases again. Note that in the range of $F_{a}$ where ordered
superlattice states can be observed ($\Psi\simeq1$), the rotation
speeds $\bar{\omega}$ are also nearly the largest. Such nearly synchronized
rotation of the DCCs introduces global dynamic chirality of the whole
system, which may be measured conveniently by an order parameter $\chi=N_{a}^{-1}\sum_{i=1}^{N_{a}}\overline{\varphi_i}$,
where $\varphi_i$ is 1(-1) if AP-$i$ rotates anti-clockwise (clockwise)
and the overbar again denotes averaging over time. Fig.\ref{Fig_Theta0.5}(f) shows the variation of $\chi$
with $F_{a}$ (right axis), where a non-monotonic behavior is present
too. With the increase of active force, $\chi$ decreases from nearly
zero at $F_{a}=0$ and then approaches a platform of $\chi\approx-1$
representing that all active particles are rotating in the same clockwise
direction, and finally increases again to $\chi\sim0$ at larger values
of $F_{a}$. In addition, the region of $F_{a}$ wherein the dynamic
chirality $\left|\chi\right|$ is maximal also coincides with
that for the optimal long-range order $\Psi$.

The above findings demonstrate a nontrivial emergence
of superlattice of DCCs driven by particle activity. In addition,
it shows that there is an optimal level of active force, where
the system is most ordered in the long range and has largest dynamic
chirality. To qualitatively understand this, we first note that a
nonzero angle $\theta$ plays an essential role in the system's dynamics.
For an isolated pair of APs with $\theta=\pi/2$, the stable
configuration should be a face-to-face one if active force is absent, as indicated in Fig.\ref{Fig_Mechanism}(a). However, if active force $F_{a}$ is not
zero, it would lead to tangential motions of both APs since $\mathbf{n}_{i}\perp\mathbf{q}_{i}$
and cause instability of the AP-pair. If $F_{a}$ is large enough,
the active force may overcome the attraction between the two APs and
the pair will break up(Fig.\ref{Fig_Mechanism}(b)). Clearly, this activity-induced-instability
of AP-pairs would not take effect if $\theta=0$ since then the active
force tends to push the two particles together. On the other hand,
for a pair of AP and PP, the active force along $\mathbf{n}$ will
also lead to tangential motion of the AP. Nevertheless, the attractive
interaction between the AP and PP is strong which tends to keep the
AP-PP pair contacted. Given that the active force is not too large,
the AP will rotate around the PP clockwise (note $\theta=\pi/2$)
as a consequence(Fig.\ref{Fig_Mechanism}(c)). And if the value of $F_{a}$ is higher, the
PP cannot attract these APs, and the DCC will also break finally(Fig.\ref{Fig_Mechanism}(d)).
Surely these rotation motion would not happen either
if the angle $\theta$ is zero or the active force $F_a=0$. What's more, the AP will rotate clockwise for $\theta\in\text{\ensuremath{\left(0,\pi\right)}}$ and anticlockwise
for $\theta\in\left(-\pi,0\right)$, respectively.

\begin{figure}
\begin{centering}
\includegraphics[width=1.0\columnwidth]{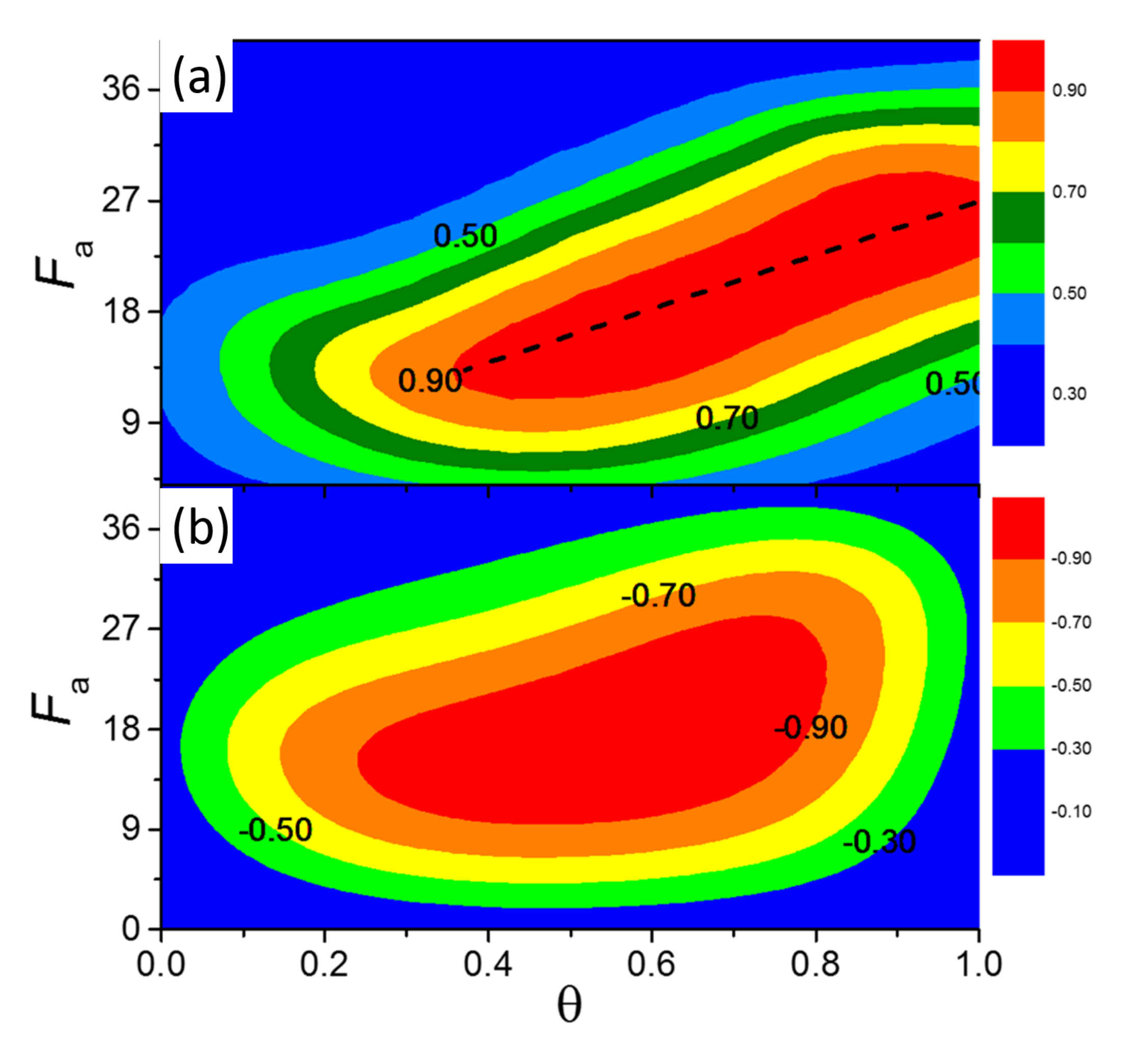}
\end{centering}
\caption{Contour plots of (a) the global ordering $\Psi$ and (b) dynamical chirality $\chi$ on the $\theta-F_{a}$
plane. The dash line in (a) represents the value of $F_a$ for maximal $\Psi$ with different fixed value of $\theta$. \label{Fig_PhaseDiagram}
}
\end{figure}

Once a DCC is formed, it will keep stable unless
the peripheral APs leave it. This allows us to estimate the onset
of instability of a DCC. Consider an AP rotating with angular speed
$\bar{\omega}$, the centripetal force required is given by $F_{\text{cen}}\sim\bar{\omega}^{2}\sigma$.
The net attractive force $F_{\text{attrac}}$ between the AP and PP
is provided by the gradient of Yukawa potential subtracted by the
partial component of active force along the direction of $-\mathbf{q}$.
Using Eq.(\ref{eq:U_AN}) with $C'=2C$ and $r_{ij}=\sigma$, we
have $F_{\text{attrac}}\sim\frac{2C}{\sigma}\left(\lambda+\sigma^{-1}\right)+F_{a}\cos\theta$.
For the DCC to be stable, it requires that $F_{\text{attrac}}\apprge F_{\text{cen}}$.
From Fig.\ref{Fig_Theta0.5}(f), one can see that $\bar{\omega}$
nearly increases linearly with $F_{a}$ for $\theta=\pi/2$ before
it reaches the maximum value, after which the ordered superlattice
lose stability. Therefore, $F_{\text{cen}}$ roughly scales as $F_{a}^{2}$,
while $F_{\text{attract}}$ increases linearly with $F_{a}$. Consequently,
$F_{\text{cen}}$ will certainly become larger than $F_{\text{attrac}}$
if $F_{a}$ becomes too strong, leading to the instability of DCC.
Surely this analysis is simple and ignores the mutual interaction
among the APs within a DCC, nevertheless, it provides a reasonable
understanding that the DCCs will become unstable if $F_{a}$ exceeds
some critical value, in accordance with the observations in Fig.(\ref{Fig_Theta0.5}).
In short, due to the deviation between the two directions $\mathbf{n}$
and $\mathbf{q}$, an appropriate level of active force can drive
the formation of DCCs, while a large active force will break them up.

To get a global picture, the contour plot of $\Psi$ in $F_{a}-\theta$ plane
is drawn in Fig.\ref{Fig_PhaseDiagram}(a). The ordered superlattice state with $\Psi\gtrsim0.9$ appears only when
both $F_{a}$ and $\theta$ are within a certain region, as shown
in red color. For larger $\theta$, it requires larger $F_{a}$ to
reach the superlattice state and also relatively larger $F_{a}$ to
destroy it. It is interesting to note that such superlattice state
can even be observed for $\theta=\pi$, while all the DCCs no longer rotate. In Fig.\ref{Fig_PhaseDiagram}(b),
the contour plot of $\chi$ is also shown in the plane of $(F_{a},\theta)$,
wherein one can easily see that an optimal level of active force with
appropriate range of biased angle $\theta$ are most favorable for
the global dynamics chirality of the system. Note that for $\theta=-\theta_{0}$,
$\Psi$ is the same as for $\theta=\theta_{0}$, while $\chi$ has
the opposite value($\chi<0$ for $\theta>0$ and $\chi>0$ for $\theta<0$).

In conclusion, we have systematically investigated
the dynamic self-assembly of a mixture of biased-active and passive
particles. The peculiar biased character of the active component,
namely, the active force is exerted along a different direction from
the orientation of anisotropic interaction, leads to the emergence
of a remarkable living superlattice with global dynamic chirality and highly ordered hexagonal structure in
both short and long ranges. Such a superlattice
state cannot be observed in the absence of active force or biased
angle, demonstrating the nontrivial roles of the both factors. To synthesize the biased-active particle experimentally, we suggest a protocol that an isotropic particle can be coated by two different layers of materials separately with the aimed biased angle, one of which produces anisotropic interaction and the other provides self-propulsion. Therefore, we believe that the biased-active particle model provides a conceptually new approach to design smart self-assembled structures, and our work may inspire a variety of following theoretical and experimental investigations in future.

\begin{acknowledgments}
This work is supported by the Ministry of Science and Technology of China(Grant Nos. 2016YFA0400904), by National Science Foundation of China (Grant Nos. 21673212, 21521001, 21473165, 21403204), and by the Fundamental Research Funds for the Central Universities (Grant Nos. WK2030020028, 2340000074).
\end{acknowledgments}


\end{document}